\documentclass[useAMS,usenatbib]{mn2e}
\usepackage{graphicx}

\title[Kinetic content of ellipsoidal configurations]{On determining the kinetic content of ellipsoidal configurations}
\author[H. Rodrigues]{Hil{\'a}rio Rodrigues\thanks{E-mail: harg@cefet-rj.br} \\
Centro Federal de Educa\c{c}\~ao Tecnol\'ogica Celso Suckow da Fonseca \\
Departamento de F\'isica\\
Av Maracan\~a 229, 20271-110, Rio de Janeiro, RJ, Brazil}

\begin{document}

\date{}

\pagerange{\pageref{firstpage}--\pageref{lastpage}} \pubyear{2012}

\maketitle

\label{firstpage}

\begin{abstract}

Determining the velocity field of structures such as galaxies, stars, and fluid planets is a relevant topic in astrophysics and astronomy. Depending on the shape of the astrophysical object, the internal velocity field may be obtained by means of analytical methods.  Specifically, ellipsoidal configurations are the most simple and natural generalization of spherically symmetric mass distributions, when rotation is present.  In this work one obtains closed analytical expressions of the velocity field and of the kinetic energy of uniform ellipsoidal configurations composed of compressible fluids. With this aim, the equation of continuity is solved allowing a description of the irrotational velocity field within the system. This permits obtaining analytical expressions of the kinetic energy for each specific mass distribution.

\end{abstract}

\begin{keywords}
celestial mechanics; stellar dynamics; ellipsoids of revolution; homoeoids; focaloids; kinetic energy.
\end{keywords}

\section{Introduction}\label{sec:Intro}

 The matter distribution within the majority of astronomical objects can not be properly described by spherically symmetric density profiles. The reason is that rotation tends to break out the spherical symmetry of such objects, explaining why ellipsoidal-like distributions of matter are common among celestial bodies e.g. satellites, planets, stars, elliptical galaxies, and clusters of galaxies  \citep{Jeans,Perek,Roberts,Chandra,Caimmia,Rasioa,Koganb,Caimmid}.  

 A historical discussion about the general equations describing the hydrostatic equilibrium of self-gravitating fluid ellipsoids is presented in the well known book of S. Chandrasekhar published in 1969 \citep{Chandra}, where the seminal works of Maclaurin, Jacobi,
Mayer, Liouville, Dirichlet, Dedekind, Riemann, and others are reviewed by the author by using a tensor virial formalism. Most of  these works provide analytical expressions of the Newtonian gravitational potential and the gravitational energy of ellipsoidal configurations. 

Such studies concerning ellipsoidal bodies have been applied during the last decades to the study of galaxies, halos, rotating stars and fluid planets. Typically, these models usually describe the internal motion of fluid structures using velocity fields with linear dependence on the rectangular coordinates \citep{Chandra,Rasioa,Rasiob,Kogana,Koganb}. However, linear velocity field  can be considered a valid approximation for rigidly rotating spheroids or ellipsoids, but not for homogeneous ellipsoidal layers, as it is shown in the present article.

 An important result concerning the study of  compressible, self-gravitating, rotating distributions of matter is the Hamy's theorem, according to which an inhomogeneous mass of fluid rotating about a fixed axis with a constant angular velocity cannot have an ellipsoidal density-distribution \citep{Hamy}. This theorem can be extended to the case of compressible fluid-masses, where the pressure is a function of the density alone, rotating under the same conditions. 

The same result is explored e.g. by Chambat,  where the author discusses the equilibrium conditions of Dedekind ellipsoids, establishing that such objects cannot be in equilibrium if rigidly rotating \citep{Chambat}. Also, Trehan and Singh discuss the equilibrium of ellipsoids under the action of uniform magnetic fields, concluding that such configurations cannot be in equilibrium, unless differential rotation is present \citep{Trehan}. More recently, Esteban and Vazquez have modeled the Earth as a system of three spheroidal shells, and show that such system cannot rotate rigidly \citep{Esteban}.    

  In addition, we recall that the gravitational potential inside inhomogeneous ellipsoidal configurations is not quadratic in the coordinates, which implies ellipsoidal shape cannot be preserved in presence of gravitational field, as shown for some heterogeneous equivalents of ellipsoidal Dedekind figures \citep{Chambat}.

The aspects addressed above entail that the true shape of celestial bodies is expected to be only slightly different from ellipsoids, even if celestial bodies in general are not rigidly rotating as shown by observations. For this reason several attempts have been made where equilibrium configurations of celestial bodies are approximated as ellipsoidal distributions of matter.

For the sake of illustration, it is worth mentioning that equations describing the dynamics of celestial bodies treated as ellipsoidal structures have been used to model {\it e.g.} the radiation of gravitational waves during the collapse of the core of a massive star, treated as a uniformly rotating homogeneous ellipsoid as in the works by Saenz and Shapiro \citep{Shapiroa,Shapirob}. 

Analogously, Lai et al use a similar set of equations to describe the hydrostatic equilibrium of both isolated or binary ellipsoidal configurations composed of compressible, differentially rotating, self-gravitating Newtonian polytropes \citep{Rasioa,Rasiob,Rasioc}. Bisnovatyi-Kogan derives a set of equations in order to describe the collapse of non-collisional dark matter and the formation of pancake universe structures treated as a uniform ellipsoidal configuration \citep{Kogana,Koganb}. However, the above quoted works in general deal with spheroids and ellipsoids, providing expressions of the Newtonian gravitational energy as well as the kinetic energy derived from velocity fields which have linear dependence on the rectangular coordinates.

In contrast, the calculation of the gravitational potential of ellipsoidal shells, {\it i.e.} systems bounded by ellipsoidal surfaces, is somewhat more difficult to perform. Fortunately, there are several works in the literature dealing with the calculation of the gravitational energy or the gravitational potential of ellipsoidal configurations in the context of Newtonian gravitation  \citep{Caimmia,Caimmib,Caimmic,Caimmid,Chambat,Neutsch,Perek,Roberts}.

 Similarly, determining the velocity field of ellipsoidal shells appears to be a rather complex problem to  be solved analytically, since for such  configurations the linear velocity field is not a full solution of the equation of continuity.

In this context, determining the kinetic content of ellipsoidal structures more complex than homogeneous spheroids or triaxial ellipsoids may be relevant for classes of celestial bodies as, for instance, Earth-like planets, neutron stars and quark stars which are assumed to be made of a rigid crust enclosing a fluid core, or a deep ocean between a rigid crust and a rigid core. Likewise   Europa-like satellites should be mentioned as they probably host a deep ocean between a rigid crust and a rigid core. Also, the description of the formation of rotating galaxies, fluid planets and stars, as well as the collapse of the supernova core when approximated by ellipsoidal configurations needs the determination of velocity field of the internal motion.

 In this work we search for closed analytical forms of the kinetic content of rotating fluid bodies of several ellipsoidal shapes. With this aim, the equation of continuity is explicitly solved, allowing analytical expressions of the irrotational velocity field inside ellipsoidal configurations. The obtained expressions of the velocity fields have non linear dependence on the rectangular coordinates.  By means of the obtained velocity fields, we calculate analytical expressions of the kinetic energy of each specific mass distribution.

The work is organized as follows: in Section \ref{sec:Sec2} one obtains the analytical expression of the kinetic energy of homogeneous fluid ellipsoids. Section \ref{sec:Sec3} provides the kinetic energy of the homogeneous homoeoid as a function of the semi-axes and the velocities of the semi-axes of the homoeoid.  In Section \ref{sec:Sec4} one  obtains the velocity field of homogeneous ellipsoidal shells of arbitrary shape, and  the kinetic energy is expressed analytically. Section \ref{sec:Sec5} is devoted to the special case of compressible focaloids, where the shells are bounded by confocal ellipsoids.  Section \ref{sec:Sec6} summarizes the main obtained results, presenting the final conclusions and comments.

\section{Compressible Ellipsoids}\label{sec:Sec2}

To begin, we shall assume a homogeneous ellipsoid of mass $M$ and semi-axes $a_{1}, a_{2}$ and $a_{3}$, composed
of a perfect and compressible fluid, rotating with angular velocity ${\bf \Omega} = (\Omega_1,\Omega_2,\Omega_3)$ relative to a fixed inertial frame of reference. Also assume that in the rotating frame formed by the principal axes of the ellipsoid $(x_{1},x_{2},x_{3})$, the fluid motion has a uniform vorticity ${\vec \xi} = (\xi_1,\xi_2,\xi_3)$. If the ellipsoid is taken to be homogeneous, the average density coincides with the local density. The average density of the ellipsoid reads
\begin{equation}
\rho=\frac{3M}{4\pi a_{1}a_{2}a_{3}}. \label{1}%
\end{equation}
   The equation describing the boundary surface of the ellipsoid in the rotating frame is expressed by
\begin{equation}
S = \frac{x_{1}^{2}}{a_{1}^{2}}+\frac{x_{2}^{2}}{a_{2}^{2}}+\frac{x_{3}^{2}}%
{a_{3}^{2}}=1. \label{0}%
\end{equation}

In the rotating frame the fluid motion inside the ellipsoid obeys the equation of continuity 
\begin{equation}
\frac{\partial {\rm v}_{1}}{\partial x_{1}}+\frac{\partial {\rm v}_{2}}{\partial x_{2}}+
\frac{\partial {\rm v}_{3}}{\partial x_{3}}=-\frac{\dot{\rho}}{\rho }, \label{2}%
\end{equation}
where {\bf v} = $({\rm v}_1,{\rm v}_2,{\rm v}_3)$ is the fluid velocity at the internal point ($x_1,x_2,x_3$) relative to the rotating frame. And from equation (\ref{1}) we have 
\begin{equation}
\frac{\dot{\rho}}{\rho}=-\sum_{i=1}^{3}\frac{\dot{a}_{i}}{a_{i}}. \label{3}%
\end{equation}

In the rotating frame, the velocity field which satisfies equation (\ref{2}) can be expressed in the form
\begin{equation}
{\bf v}=\sum_{i=1}^{3} {\rm v}_{i} \hat{e}_{i}, \label{4}
\end{equation}
with each component expressed as a linear function of the rectangular coordinates, which are given by
\begin{equation}
{\rm v}_{i}=\frac{\dot{a}_{i}}{a_{i}}x_{i} +  \frac{a_i}{a_j} \Lambda_k  x_j -  \frac{a_i}{a_k}\Lambda_j x_{k} , \label{4a}
\end{equation}
with $i \ne j \ne k $. 

 The quantities $\Lambda_i$ ($i=1, 2, 3$) present in equation (\ref{4a}) are given by
\begin{equation}
\Lambda_{i} = - \frac{a_{j} a_{k}}{a_{j}^{2}+a_{k}^{2}}\xi_{i}, \, \, \, (i \neq j \neq k), \label{eqLambda}
\end{equation}
where the terms $\xi_i$ are the components of the vorticity which is defined to be the curl of the velocity, ${\bf \xi} = \Delta \times   {\bf v}$, which are dimensioned as the inverse of time. The same velocity field given by equation (\ref{4a}) is derived by Chandrasekhar \citep{Chandra}, but using a quite different approach. 

It is important to mention that the velocity field given by equation (\ref{4a}) preserves the ellipsoidal boundary of the body. To show this, consider the time derivative of the surface equation given in equation (\ref{0}). This yields 
\begin{equation}
\frac{1}{2}\frac{dS}{dt}=\sum_{i=1}^{3}\left( \frac{x_{i}}{a_{i}^{2}}{\rm v}_{i}-
\frac{x_{i}^{2}}{a_{i}^{2}}\frac{\dot{a}_{i}}{a_{i}}\right) \label{eqsurf}.
\end{equation}
Inserting (\ref{4}) in the right-hand side of equation (\ref{eqsurf}), we get
\begin{equation} 
\frac{dS}{dt} =0  \label{eqsurf2},
\end{equation}
showing that the ellipsoidal boundary given by equation (\ref{0}) is maintained every time during the fluid motion.

The components of the velocity field in the inertial frame, {\bf V}, resolved along the coordinate axes of the rotating frame, can be expressed in terms of {\bf v} by the relation
\begin{equation}
{\rm V}_{i}= {\rm v}_{i} +  \varepsilon_{ijk} \Omega_j x_{k} ,  \label{4aa}
\end{equation}
where $\varepsilon_{ijk}$ is the Levi-Civita symbol, which is $+1$ if $(i,j,k)$ is an even permutation of $%
(1,2,3)$, $-1$ if it is an odd permutation, and $0$ for any repeated index.  The summation convention is adopted in equation (\ref{4aa}). Thus, we obtain explicitly 
\begin{equation}
{\rm V}_{i} = {\rm v}_i  + \Omega_j x_{k} - \Omega_k x_j , \, \, \, \, (i \ne j \ne k ) , \label{4aa2}
\end{equation}
with the components ${\rm v}_i$ given by equation (\ref{4a}).
 
The kinetic energy associated with the motion of the content of the ellipsoid can be obtained by the integral
\begin{equation}
T=\int \frac{1}{2} \rho \left|{\bf V}\right|^{2} dx_1 dx_2 dx_3. \label{8}%
\end{equation}

Carrying out the integration on the entire volume of the body, we find out the following simple quadratic form of the kinetic energy of the ellipsoid:
\begin{eqnarray}
T&=&\frac{1}{10}M \left\{ \sum_{i=1}^{3} \dot{a}_{i}^{2} - 4 \sum_{i \ne j \ne k} a_i a_j \Lambda_k \Omega_k \right. \nonumber \\
&& + \left.\sum_{i \ne j \ne k}(a_i^2 + a_j^2) \left( \Lambda_k^2 + \Omega_k^2 \right) \right\} .  \label{9a}
\end{eqnarray}

This is the same expression given by Chandrasekhar \citep{Chandra}. The first term on the right-hand side of the equation (\ref{9a}) accounts for the kinetic energy associated with the fluid radial motion of the ellipsoid. The second and third terms come from the vorticity motions and the rotation of the fluid, providing the total rotational energy of the system.

\section{Compressible Homoeoids}\label{sec:Sec3}

We shall now consider a homogeneous ellipsoidal shell of density $\rho$ which
is bounded internally by the ellipsoidal surface of semi-axes $A_{i}%
$\ ($i=1,2,3$) and externally by a concentric ellipsoidal surface of semi-axes
$a_{i}$, being $A_{i}=ma_{i}\ $($0\leq m<1$). This form of mass distribution defines a special class of ellipsoidal shells, the {\it homoeoids}. 

In the rotating frame, the equation describing an ellipsoidal surface within the homoeoid, including the outer and the inner boundary surfaces, is given by
\begin{equation}
 \frac{x_{1}^{2}}{u^2 a_{1}^{2}}+\frac{x_{2}^{2}}{u^2 a_{2}^{2}}+\frac{x_{3}^{2}}%
{u^2 a_{3}^{2}}=1,   \label{0b}
\end{equation} 
where we have $m\leq u\leq 1$.

 The mass density of the homogeneous homoeoid reads
\begin{equation}
\rho=\frac{3M}{4\pi a_{1}a_{2}a_{3}(1-m^{3})},\label{14}%
\end{equation}
where $M$ is the mass of the homoeoid. Thus, the time derivative of the density is found to be 
\begin{equation}
\frac{\dot{\rho}}{\rho}=-\left(  \frac{\dot{a}_{1}}{a_{1}}+\frac{\dot{a}_{2}%
}{a_{2}}+\frac{\dot{a}_{3}}{a_{3}} \right) + \frac{3m^{3}}{1-m^{3}}\frac{\dot{m}}{m} 
,\label{15}%
\end{equation}
where $\dot{m}$ can be obtained from
\begin{equation}
\frac{\dot{m}}{m}=\frac{\dot{A}_{i}}{A_{i}} - \frac{\dot{a}_{i}}{a_{i}%
}.\label{16}%
\end{equation}

Equation (\ref{15}) is to be substituted in the right-hand side of equation (\ref{2}). In this case the velocity field which satisfies the equation of continuity, calculated in the rotating frame, has a quite more complex form, and whose components are explicitly given by
\begin{eqnarray}
{\rm v}_i&=&\left[\frac{\dot{a}_{i}}{a_{i}}-\frac{m^{3}}{1-m^{3}}\left(1-\frac{1}{u^{3}}\right) \frac{\dot{m}}{m}\right] x_{i} \nonumber \\
&+&  \frac{a_i}{a_j} \Lambda_k x_j - \frac{a_i}{a_k}\Lambda_jx_{k} ,\label{17a}
\end{eqnarray}
where $i \ne j \ne k $, and $m \le u \le 1$ according to equation (\ref{0b}). 

The velocity field expressed by equation (\ref{17a}) preserves the ellipsoidal boundaries of the body. In fact, for any point at the outer boundary surface of the homoeoid, for which $u=1$, it assumes the simplified linear form
\begin{equation}
{\rm v}_i^{\rm out} =  \frac{\dot{a}_{i}}{a_{i}} x_{i} 
+  \frac{a_i}{a_j} \Lambda_k x_j - \frac{a_i}{a_k}\Lambda_jx_{k} . \label{17a1}%
\end{equation}

We can see that at any point of the inner boundary surface of the homoeoid, for which $u=m$, we have also
\begin{equation}
{\rm v}_i^{\rm in} =  \frac{\dot{A}_{i}}{A_{i}} x_{i} 
+  \frac{A_i}{A_j} \Lambda_k  x_j - \frac{A_i}{A_k}\Lambda_j x_{k} . \label{17a2}%
\end{equation}

Thus, according to the arguments present in Section \ref{sec:Sec2}, one shows that the velocity field expressed by equation (\ref{17a}) preserves the inner and the outer ellipsoidal boundaries every time, because it satisfies equations similar to (\ref{eqsurf}) and (\ref{eqsurf2}). Interestingly, substituting $m = 0$ in equation (\ref{17a}) we recover the velocity field (\ref{4a}), valid for triaxial ellipsoids. 

The components of the velocity field in the inertial frame are written by using the same prescription given in equation (\ref{4aa}). In obtaining the kinetic energy of the homoeoid, we have to carry out the integral defined in equation (\ref{8}). To achieve it, we map the homoeoid into a spherical shell where the dimensionless radial coordinate $r=u$ varies from $r=m$ to $r=1$. To get this, consider the following variables transformations:
\begin{equation}
x_1 = a_1 r \sin\theta \cos\phi,  
\end{equation}
\begin{equation}
x_2 = a_2 r \sin\theta \sin\phi,   
\end{equation}
and
\begin{equation}
x_3 = a_3 r \cos\theta ,  
\end{equation}
with $\theta$ and $\phi$ being the polar spherical coordinates. After transforming the integral (\ref{8}) by using the mapping described above, we arrive at the following integral of the kinetic energy of the homoeoid: 
\begin{equation}
T= \frac{1}{2} \rho a_{1}a_{2}a_{3}\int_{0}^{2\pi }\int_{0}^{\pi
}\int_{m}^{1} \left| {\bf V} \right|^{2}r^{2}\sin\theta dr d\theta d\phi \label{Kin22}.
\end{equation}
where the velocity field $\bf V$ is defined by equation (\ref{4aa}). Now, inserting equation (\ref{17a}) into equation (\ref{4aa}), and the equation (\ref{14}) into equation (\ref{Kin22}), and carrying out the triple integration, we obtain the following expression for the total kinetic energy of the homoeoid: 
\begin{eqnarray}
\frac{T}{\left(M/10\right)} &=&  f(m) \left( \dot{a}_{1}^{2}+\dot{a}_{2}^{2}+\dot{a}
_{3}^{2} \right)   \nonumber \\
&& + \ g(m)\left( a_{1}\dot{a}_{1}+a_{2}\dot{a}_{2}+a_{2}\dot{a}_{2}\right) \dot{
m} \nonumber \\
&& + \ h(m)\left( a_{1}^{2}+a_{2}^{2}+a_{3}^{2}\right) \dot{m}^{2}  \nonumber  \\
&& + \ f(m) \sum_{i \ne j \ne k}(a_i^2 + a_j^2)(\Lambda_k^2+\Omega_k^2)  \nonumber \\
&& -4 \ f(m)\sum_{i \ne j \ne k} a_i a_j \Lambda_k \Omega_k . \label{9}   
\end{eqnarray}
where
\begin{equation}
f(m)=\frac{m^{4}+m^{3}+m^{2}+m+1}{m^{2}+m+1}, \label{eq27b}
\end{equation}%
\begin{equation}
g(m)=\frac{m^{2}(2m^{3}+4m^{2}+6m+3)}{\left( m^{2}+m+1\right) ^{2}}, 
\end{equation}%
and
\begin{equation}
h(m)=\frac{m^{3}\left( m^{3}+3m^{2}+6m+5\right) }{\left( m^{2}+m+1\right)
^{3}} . \label{eq29b} 
\end{equation}

 Notice that in the limit $A_i \rightarrow 0$, which implies $m \rightarrow 0$, equation (\ref{9}) reduces to equation (\ref{9a}).  Also, the infinitely thin homoeoid is placed on the external boundary as $m \rightarrow 1$, which implies $\dot{m} \rightarrow 0$. Assuming the additional conditions $\Lambda_i \rightarrow 0$ (null vorticity on the boundary or within an infinitely thin homoeoid), and so  $\dot{\Lambda_i} \rightarrow 0$, the kinetic energy $dT$ of an infinitely thin homoeoid of infinitesimal mass $d M$ is expressed as
\begin{equation}
dT=\frac{1}{6}dM \left\{ \sum_{i=1}^{3} \dot{a}_{i}^{2} + \sum_{i \ne j \ne k}(a_i^2 + a_j^2)  \Omega_k^2 \right\} .  \label{9a23}
\end{equation}

 Finally, there is an interesting limit to be mentioned, which is related to the special case of non rotating spherical shell configurations, for which we define $A_1=A_2=A_3=A$, $a_1=a_2=a_3=a$. Assuming so $\Omega = 0$ (null rotation) and $\Lambda=0$ (null vorticity), we get from equation (\ref{17a}) the following expression for the velocity field
\begin{equation}
{\rm v}(r) = \frac{ a^2 \dot{a} - A^2\dot{A}  }{a^3 - A^3} \left(r - \frac{A^3}{r^2} \right) + \frac{A^2}{r^2} \dot{A} , \label{esfv}
\end{equation}
where here $r$ stands for the radial distance from the centre of the spherical shell, which has the dimension of length. From equation (\ref{9}) one obtains the kinetic energy  
\begin{equation}
T =\frac{3}{10} M \left( t_{11} \dot{a}^{2} + t_{22} \dot{A}^{2} + t_{12} \dot{a}\dot{A}
\right) , \label{K2}
\end{equation}
where
\begin{equation}
t_{11} =  \frac{5m^{3}+6m^{2}+3m +1} {(m^{2}+m+1)^3}, \label{tesf11} 
\end{equation}%
\begin{equation}
t_{12} = 3 m^2 \frac{m^{2}+3m +1} {(m^{2}+m+1)^3},  \label{tesf12}
\end{equation}
and
\begin{equation} 
t_{22} = h(m), \label{t22esf}
\end{equation}
with $h(m)$ given by equation (\ref{eq29b}), and $m$ being defined by $m=\frac{A}{a}$, with $0< m < 1$. The same expression is obtained by substituting the velocity field (\ref{esfv}) directly into equation (\ref{Kin22}) and then carrying out the integration \citep{Rodrigues}. 

Notice that the limit $x \rightarrow 1$, implies the infinitely thin spherical shell is placed on the external boundary, from which we get $A \rightarrow a$, $\dot{A} \rightarrow \dot{a}$, $t_{11} \rightarrow \frac{5}{9}$, $ t_{12} \rightarrow \frac{5}{9}$, $t_{22} \rightarrow \frac{5}{9}$, and so equation (\ref{K2}) converges to the following expression for the differential kinetic energy:
\begin{equation}
dT=\frac{1}{2}dM \dot{a}^{2} .  \label{K2b}
\end{equation}
 
We further note that equation (\ref{K2b}) could be obtained directly from equation (\ref{9a23}) in the limit $\dot{a}_1 \rightarrow \dot{a}$, $\dot{a}_2 \rightarrow \dot{a}$, $\dot{a}_3 \rightarrow \dot{a}$, and  $\Omega_i = 0$.

\section{Ellipsoidal Systems of Arbitrary Shape}\label{sec:Sec4}

In this section we consider a homogeneous ellipsoidal layer of mass $M$ which
is bounded internally by the ellipsoidal surface of semi-axes $A_{i}$\ ($i=1,2,3$) and externally by the ellipsoidal surface of semi-axes
$a_{i}$. For this general configuration the ratios between the parallel semi-axes define the parameters $l$, $m$ and $n$, given by: $l = \frac{A_1}{a_1}$, $m = \frac{A_2}{a_2}$, and $n=\frac{A_3}{a_3}$.  

In the rotating frame, the equation describing an ellipsoidal surface within the layer, including the outer and the inner boundary surfaces, has the form
\begin{equation}
 \frac{x_{1}^{2}}{\lambda^2 a_{1}^{2}}+\frac{x_{2}^{2}}{\mu^2 a_{2}^{2}}+\frac{x_{3}^{2}}%
{\nu^2 a_{3}^{2}}=1,   \label{0b2}
\end{equation} 
where $l\leq \lambda \leq 1$, $m\leq \mu \leq 1$, and $n\leq \nu \leq 1$.

 The mass density of the homogeneous shell thus reads
\begin{equation}
\rho=\frac{3M}{4\pi a_{1}a_{2}a_{3}(1-l m n)}, \label{14c}%
\end{equation}
and so the time derivative of the density is given by 
\begin{equation}
\frac{\dot{\rho}}{\rho}=-\left(  \frac{\dot{a}_{1}}{a_{1}}+\frac{\dot{a}_{2}%
}{a_{2}}+\frac{\dot{a}_{3}}{a_{3}} \right) + \frac{lmn}{1-lmn} \left( \frac{\dot{l}}{l} + \frac{\dot{m}}{m} +\frac{\dot{n}}{n} \right)
,\label{15c}%
\end{equation}
where
\begin{equation}
\frac{\dot{l}}{l}=\frac{\dot{A}_{1}}{A_{1}}-\frac{\dot{a}_{1}}{a_{1}},\label{16c1}%
\end{equation}
\begin{equation}
\frac{\dot{m}}{m}=\frac{\dot{A}_{2}}{A_{2}}-\frac{\dot{a}_{2}}{a_{2}},\label{16c2}%
\end{equation}
and
\begin{equation}
\frac{\dot{n}}{n}=\frac{\dot{A}_{3}}{A_{3}}-\frac{\dot{a}_{3}}{a_{3}}.\label{16c3}%
\end{equation}

 The velocity field which satisfies the equation of continuity (\ref{2}), determined in the rotating frame, is given by the following Cartesian components
\begin{eqnarray}
{\rm v}_i&=&\left[\frac{\dot{a}_{i}}{a_{i}}-\frac{lmn}{3(1-lmn)}\left(1-\frac{1}{u^{3} }\right) \left( \frac{\dot{l}}{l} + \frac{\dot{m}}{m} +\frac{\dot{n}}{n} \right)  \right] x_{i} \nonumber \\
&+&  C_i \left(\frac{1-\lambda}{1-l}\right) \left(\frac{1-\mu}{1-m}\right) \left(\frac{1-\nu}{1-n}\right)x_i    \nonumber \\ 
&+&  \frac{{\tilde a}_i}{{\tilde a}_j} \Lambda_k x_j - \frac{{\tilde a}_i}{{\tilde a}_k}\Lambda_j x_{k}, \ \ \ \ \ i \ne j \ne k  ,\label{17ac1}
\end{eqnarray}
where ${\tilde a}_1 = \lambda a_1$, ${\tilde a}_2 = \mu a_2$, and ${\tilde a}_3 = \nu a_3$.

The parameters $C_i$ in equation (\ref{17ac1}) are given by
\begin{equation}
C_{1}= \frac{1}{3}\left(\frac{2\dot{l}}{l} - \frac{\dot{m}}{m} - \frac{\dot{n}}{n}%
\right) , 
\end{equation}
\begin{equation}
C_{2}=\frac{1}{3}\left(\frac{2\dot{m}}{m} - \frac{\dot{l}}{l} - \frac{\dot{n}}{n}%
\right) , 
\end{equation}
and
\begin{equation}
C_{3}= \frac{1}{3}\left(\frac{2\dot{n}}{n} - \frac{\dot{l}}{l} - \frac{\dot{m}}{m}%
\right) ,
\end{equation}
with $(l m n)^{1/3} \le u \le 1$.
 
 On the external surface of the ellipsoidal layer we have $u = \lambda =\mu = \nu = 1$, and thus each component of the velocity field given by equation (\ref{17ac1}) takes the same linear form given by equation (\ref{4a}). Furthermore, on the inner boundary surface, where $u=(l m n)^{1/3}$, the velocity field assumes the linear form given by equation (\ref{17a2}). Thus, the obtained velocity field preserves the form of the ellipsoidal boundaries, because it satisfies equations similar to equations (\ref{eqsurf}) and (\ref{eqsurf2}), as required. It is worth noting that for $l=m=n$, and ${\dot l} = {\dot m} = {\dot n} $, equation (\ref{17ac1}) recovers the velocity field (\ref{17a}), valid for homogeneous triaxial homoeoids.

In order to obtain the components of the velocity field in the inertial frame, we use again the prescription given by equation (\ref{4aa}). The  kinetic energy associated with the motion of the fluid is obtained by carrying out the integral defined by equation (\ref{8}). Adopting the same variable transformations used in Section \ref{sec:Sec3}, and substituting equations (\ref{14c}) and (\ref{17ac1}) into equation (\ref{8}), one obtains 
\begin{eqnarray}
\frac{T}{\left(M/10\right)} &=& f_0 \left( \dot{a}_{1}^{2}+\dot{a}_{2}^{2}+\dot{a}
_{3}^{2}\right)  \nonumber \\
&& +  \left[ g_1 a_{1}\dot{a}_{1}+ g_2 \left(a_{2}\dot{a}_{2}+a_{3}\dot{a}_{3} \right) \right] \left( \frac{\dot{
l}}{l} \right) \nonumber  \\
&& +  \left[ g_1 a_{2}\dot{a}_{2}+ g_2 \left(a_{1}\dot{a}_{1}+a_{3}\dot{a}_{3} \right) \right] \left( \frac{\dot{
m}}{m} \right) \nonumber  \\
&& +  \left[ g_1 a_{3}\dot{a}_{3}+ g_2 \left(a_{1}\dot{a}_{1}+a_{2}\dot{a}_{2} \right) \right] \left( \frac{\dot{
n}}{n} \right) \nonumber  \\
&& +  \left[ h_1 \left( a_1^2 + a_2^2 \right) + h_2 a_3^2 \right] \left(\frac{\dot{l} \dot{m}}{l m} \right) \nonumber \\
&& +  \left[ h_1 \left( a_1^2 + a_3^2 \right) + h_2 a_2^2 \right] \left(\frac{\dot{l} \dot{n}}{l n} \right) \nonumber \\
&& +  \left[ h_1 \left( a_2^2 + a_3^2 \right) + h_2 a_1^2 \right] \left(\frac{\dot{m} \dot{n}}{m n} \right) \nonumber \\
&& +   \left[  h_3 a_{1}^{2} + h_4 \left( a_{2}^{2}+a_{3}^{2} \right) \right] \left(\frac{\dot{l}}{l}\right)^{2} \nonumber   \\
&& +   \left[  h_3 a_{2}^{2} + h_4 \left( a_{1}^{2}+a_{3}^{2} \right) \right] \left(\frac{\dot{m}}{m}\right)^{2} \nonumber   \\
&& +   \left[  h_3 a_{3}^{2} + h_4 \left( a_{1}^{2}+a_{2}^{2} \right) \right] \left(\frac{\dot{n}}{n}\right)^{2}  \nonumber   \\
&& +   \sum_{i \ne j \ne k} \left( f_{ik} a_i^2 + f_{jk} a_j^2 \right) \Lambda_k^2  \nonumber \\
&& +   f_0 \sum_{i \ne j \ne k} \left( a_i^2 + a_j^2 \right) \Omega_k^2   \nonumber \\
&& - 2 \sum_{i \ne j \ne k} f_i a_i a_j \Lambda_k \Omega_k , \label{Kin2} 
\end{eqnarray}
where
\begin{equation}
f_0  = \frac{(l m n)^{4/3}+lmn+(l m n)^{2/3}+(l m n)^{1/3}+1}{(l m n)^{2/3}+(l m n)^{1/3}+1} , \label{eq46} 
\end{equation}
\begin{equation}
 f_1=\frac{ 2 (l m n)^{1/3} - l^3 m n^2  - l m^3 n^2 } {(l m n)^{1/3} \left( 1- l m n \right) } , 
\end{equation}
\begin{equation}
 f_2=\frac{ 2 (l m n)^{1/3} - l^3 m^2 n  - l m^2 n^3 } {(l m n)^{1/3} \left( 1- l m n \right) } , 
\end{equation}
\begin{equation}
 f_3=\frac{ 2 (l m n)^{1/3} - l^2 m^3 n  - l^2 m n^3 } {(l m n)^{1/3} \left( 1- l m n \right) } , 
\end{equation}
\begin{equation}
g_1=  (l m n)^{2/3} \frac{  2 l^2 m^2  n^2  -3 l m n + (l m n)^{1/3} } { \left( 1 -  l m n \right)^2} , 
\end{equation}
\begin{equation}
g_2=  (l m n)^{2/3} \frac{  (l m n)^{1/3} - l m n } {\left( 1 -  l m n \right)^2} ,
\end{equation}
\begin{equation}
h_1= - (l m n)^{5/3} \frac{  2 (l m n)^{1/3} - l m n  -1 } {\left( 1- l m n \right)^3} ,
\end{equation}
\begin{equation}
h_2= -2(l m n)^{5/3} \frac{ (l m n)^{1/3}  -1 } {\left( 1- l m n \right)^3} ,
\end{equation}
\begin{equation}
h_3= -(l m n)^{5/3} \frac{ l^2 m^2 n^2 -3 l m n + (l m n)^{1/3} + 1 } {\left( 1- l m n \right)^3}  ,
\end{equation}
\begin{equation}
 h_4= \frac{1}{2} h_2 , 
\end{equation}
\begin{equation}
 f_{12}=\frac{ (l m n)^{1/3} - l^4 m^2  } {(l m n)^{1/3} \left( 1- l m n \right) } , 
\end{equation}
\begin{equation}
 f_{13}=\frac{ (l m n)^{1/3} - l^4 n^2  } {(l m n)^{1/3} \left( 1- l m n \right) } , 
\end{equation}
\begin{equation}
 f_{21}=\frac{ (l m n)^{1/3} - m^4 l^2  } {(l m n)^{1/3} \left( 1- l m n \right) } , 
\end{equation}
\begin{equation}
 f_{23}=\frac{ (l m n)^{1/3} - m^4 n^2  } {(l m n)^{1/3} \left( 1- l m n \right) } , 
\end{equation}
\begin{equation}
 f_{31}=\frac{ (l m n)^{1/3} - n^4 l^2  } {(l m n)^{1/3} \left( 1- l m n \right) } , 
\end{equation}
and
\begin{equation}
 f_{32}=\frac{ (l m n)^{1/3} - n^4 m^2  } {(l m n)^{1/3} \left( 1- l m n \right) } .  \label{f32}
\end{equation}

Notice that for $l=m=n$ and $ \dot{l}=\dot{m}=\dot{n}$, equation (\ref{Kin2}) recovers equation (\ref{9}), valid for the homogeneous homoeoid discussed in Section \ref{sec:Sec3}, since in this case equation (\ref{eq46}) reduces to equation (\ref{eq27b}). In a similar way, the infinitely thin ellipsoidal corona is placed on the external boundary as $l \rightarrow 1$, $m \rightarrow 1$, $n \rightarrow 1$, which implies $\dot{l} \rightarrow 0$, $\dot{m} \rightarrow 0$, $\dot{n} \rightarrow 0$. Assuming again the additional conditions $\Lambda_i \rightarrow 0$ and ${\dot \Lambda_i} \rightarrow 0$ (i.e. null vorticity on the boundary or within an infinitely thin ellipsoidal corona) we see that equation (\ref{Kin2}) reduces to equation (\ref{9a23}).  

We may extend the calculations presented above in order to treat any ellipsoidal configuration composed of an arbitrary number of subsystems. With this aim, we divide the system into a central ellipsoid of mass $M_0$ surrounded by $N$ ellipsoidal layers of mass $M_n$, $1\le n \le N$, under the constraint $\sum_{n=0}^N M_n =M$, where $M$ is the total mass of the system. For $n=1,\ldots, N$, the $n-$th shell is bounded by the ellipsoidal surfaces defined by the semi-axes $(a_1^{n},a_2^{n},a_3^{n})$ and  $(a_1^{n-1},a_2^{n-1},a_3^{n-1})$. For $n=0$, $a_1^0$, $a_2^0$, $a_3^0$, are the semi-axes of the central ellipsoid. Here, we define the semi-axes ratios $m_i^n =\frac {a_i^{n}}{a_i^{n-1}}$, valid for $1 \le n \le N$. 

 Thus, the kinetic energy of the compound system can be written as
\begin{equation}
T=\sum_{n=0}^{N} T_{n} , \label{82}%
\end{equation}
where $T_0$ is the kinetic energy of the central ellipsoid, and the kinetic energy of the remaining shells, $T_n$, are  determined by equation (\ref{Kin2}), with the substitutions: $M \rightarrow M_n $, $l \rightarrow m_1^n$, $m \rightarrow m_2^n$, $n \rightarrow m_3^n$, $\dot{l} \rightarrow \dot{m}_1^n$, $\dot{m} \rightarrow \dot{m}_2^n$, $\Lambda_i \rightarrow \Lambda_i^n$ and $\Omega_i \rightarrow \Omega_i^n$, into equations (\ref{Kin2}) to (\ref{f32}). Conversely, the kinetic energy of the  central ellipsoid , $T_0$, is calculated by using equation (\ref{9a}), with the substitutions: $M \rightarrow M_0$,  $\Lambda_i \rightarrow \Lambda_i^0$, and $\Omega_i \rightarrow \Omega_i^0$.

\section{Compressible Focaloids}\label{sec:Sec5}

In this section we describe the special case of focaloids, which has been
considered in astronomical subjects \citep{Perek,Roberts}.  Focaloids are shells bounded
by two concentric, confocal, triaxial ellipsoids. If the inner boundary surface is specified by
\begin{equation}
\frac{x_{1}^{2}}{A_{1}^{2}}+\frac{x_{2}^{2}}{A_{2}^{2}}+\frac{x_{3}^{2}}{%
A_{3}^{2}}=1,
\end{equation}%
with semi-axes $A_{1}$, $A_{2}$, $A_{3}$, the outer ellipsoidal surface is described by
\begin{equation}
\frac{x_{1}^{2}}{A_{1}^{2}+\lambda }+\frac{x_{2}^{2}}{A_{2}^{2}+\lambda }+%
\frac{x_{3}^{2}}{A_{3}^{2}+\lambda }=1,
\end{equation}%
with $A_{1}>A_{2}>A_{3}$, and $\lambda >0$. Confocal ellipsoids have the same foci, which for the system above are defined by
\begin{eqnarray}
f_{1}^{2} &=&A_{2}^{2}-A_{3}^{2}=\left( A_{2}^{2}+\lambda \right) -\left(
A_{3}^{2}+\lambda \right) ,  \nonumber \\
f_{2}^{2} &=&A_{1}^{2}-A_{3}^{2}=\left( A_{1}^{2}+\lambda \right) -\left(
A_{3}^{2}+\lambda \right) , \\
f_{3}^{2} &=&A_{1}^{2}-A_{2}^{2}=\left( A_{1}^{2}+\lambda \right) -\left(
A_{2}^{2}+\lambda \right) .  \nonumber
\end{eqnarray}%

A focaloid can be used as a construction element of a matter distribution. A
remarkable property is that two different, concentric, confocal focaloids of the same
mass produce the same action on a test mass in the exterior region. This result is known as the MacLaurin's theorem \citep{Chandra}. 

The density of the homogeneous focaloid of mass $M$ is given by equation (\ref{14c}), where however are valid the following relationships 
\begin{equation}
a_{i}^{2}=A_{i}^{2}+\lambda , \ \ \ \ i=1,2,3 ,
\end{equation}
between the semi-axes of the focaloid. On the other hand, the relationships $A_{1}=la_{1}$, $A_{2}=ma_{2}$, and $A_{3}=na_{3}$ still hold, provided that
\begin{equation}
l=\left( 1-\frac{\lambda }{a_{1}^{2}}\right) ^{1/2}, \label{defl}
\end{equation}%
\begin{equation}
m=\left( 1-\frac{\lambda }{a_{2}^{2}}\right) ^{1/2}, \label{defm}
\end{equation}%
and
\begin{equation}
n=\left( 1-\frac{\lambda }{a_{3}^{2}}\right) ^{1/2}. \label{defn}
\end{equation}%

Thus, the time derivative of the mass density of the focaloid is also given by
equation (\ref{15c}) where, however, we should introduce the following substitutions
\begin{eqnarray}
\dot{l} &=&\frac{\lambda}{l} \frac{1}{a_1^2} \left(\frac{\dot{a}_{1}}{a_{1}}-\frac{\dot{\lambda}}{2 \lambda}\right) , \\
\dot{m} &=& \frac{\lambda}{m} \frac{1}{a_2^2} \left(\frac{\dot{a}_{2}}{a_{2}}-\frac{\dot{\lambda}}{2 \lambda}\right) , \\
\dot{n} &=& \frac{\lambda}{n} \frac{1}{a_3^2} \left(\frac{\dot{a}_{3}}{a_{3}}-\frac{\dot{\lambda}}{2 \lambda}\right) .
\end{eqnarray}

The velocity field resolved in the rotating frame has the same form given by
equation (\ref{17ac1}).  Thus, the kinetic energy of the focaloid is obtained by means of the same procedure described in Section \ref{sec:Sec3}, which can be expressed by the quadratic form
\begin{eqnarray}
\frac{T}{\left(M/10\right)} &=&  T_{11}\dot{a}_{1}^{2}+T_{22}\dot{a}_{2}^{2}+T_{33}%
\dot{a}_{3}^{2}+T_{44}\dot{\lambda}^{2}  \nonumber \\
&&+ T_{12}\dot{a}_{1}\dot{a}_{2}+T_{13}\dot{a}_{1}\dot{a}_{3}+T_{23}\dot{a}%
_{2}\dot{a}_{3} \nonumber \\ 
&&+ \left(T_{14}\dot{a}_{1} + T_{24}\dot{a}_{2}+T_{34}\dot{a}_{3}\right)\dot{\lambda} , \nonumber \\
&& +   \sum_{i \ne j \ne k} \left( f_{ik} a_i^2 + f_{jk} a_j^2 \right) \Lambda_k^2  \nonumber \\
&& +   f_0 \sum_{i \ne j \ne k} \left( a_i^2 + a_j^2 \right) \Omega_k^2   \nonumber \\
&& - 2 \sum_{i \ne j \ne k} f_i a_i a_j \Lambda_k \Omega_k , \label{eq75}
\end{eqnarray}
where the parameters $T_{ij}$ are given by 
\begin{equation}
T_{11}=\frac{m^{2}n^{2}}{l^{2}}\left[ F_{1}a_{1}^{2}+F_{2}\left(
a_{2}^{2}+a_{3}^{2}\right) \right] \frac{\lambda ^{2}}{a_{1}^{6}}+\frac{mn}{l%
}F_{3}\frac{\lambda }{a_{1}^{2}}+f_{0}, \label{eqk11}
\end{equation}
\begin{equation}
T_{22}=\frac{l^{2}n^{2}}{m^{2}}\left[ F_{1}a_{2}^{2}+F_{2}\left(
a_{1}^{2}+a_{3}^{2}\right) \right] \frac{\lambda ^{2}}{a_{2}^{6}}+\frac{ln}{m%
}F_{3}\frac{\lambda }{a_{2}^{2}}+f_{0}, 
\end{equation}
\begin{equation}
T_{33}=\frac{l^{2}m^{2}}{n^{2}}\left[ F_{1}a_{3}^{2}+F_{2}\left(
a_{1}^{2}+a_{2}^{2}\right) \right] \frac{\lambda ^{2}}{a_{3}^{6}}+\frac{lm}{n%
}F_{3}\frac{\lambda }{a_{3}^{2}}+f_{0},
\end{equation}
\begin{eqnarray}
T_{44} &=&\frac{1}{4} m^2 \left\{ G_1\left(a_1^2+a_3^2\right) + 2F_2a_2^2 \right\} \frac{1}{a_1^2  a_3^2} \nonumber \\
&&+ \frac{1}{4} n^2 \left\{ G_1\left(a_1^2+a_2^2\right) + 2F_2a_3^2 \right\} \frac{1}{a_1^2  a_2^2} \nonumber \\
&&+\frac{1}{4} l^2 \left\{ G_1\left(a_2^2+a_3^2\right) + 2F_2a_1^2 \right\} \frac{1}{a_2^2  a_3^2} \nonumber \\
&&+\frac{1}{4} \frac{m^2 n^2}{l^2} \left\{ F_1 a_1^2 + F_2 \left(a_2^2+a_3^2\right) \right \} \frac{1}{a_1^4} \nonumber \\
&&+\frac{1}{4} \frac{l^2 n^2}{m^2} \left\{ F_1 a_2^2 + F_2 \left(a_1^2+a_3^2\right) \right \} \frac{1}{a_2^4} \nonumber \\ 
&&+\frac{1}{4} \frac{l^2 m^2}{n^2} \left\{ F_1 a_3^2 + F_2 \left(a_1^2+a_2^2\right) \right \} \frac{1}{a_3^4} ,
\end{eqnarray}
\begin{eqnarray}
T_{12} &=&n^{2}\left[ G_{1}\left( a_{1}^{2}+a_{2}^{2}\right) +2F_{2}a_{3}^{2}%
\right] \frac{\lambda ^{2}}{a_{1}^{3}a_{2}^{3}} \nonumber \\
&&+ G_{2}\left( \frac{ln}{m}
a_{1}^{4}+\frac{mn}{l}a_{2}^{4}\right) \frac{\lambda }{a_{1}^{3}a_{2}^{3}},
\end{eqnarray}
\begin{eqnarray}
T_{13} &=&m^{2}\left[ G_{1}\left( a_{1}^{2}+a_{3}^{2}\right) +2F_{2}a_{2}^{2}%
\right] \frac{\lambda ^{2}}{a_{1}^{3}a_{3}^{3}} \nonumber \\
&&+G_{2}\left( \frac{lm}{n}%
a_{1}^{4}+\frac{mn}{l}a_{3}^{4}\right) \frac{\lambda }{a_{1}^{3}a_{3}^{3}},
\end{eqnarray}
\begin{eqnarray}
T_{23} &=&l^{2}\left[ G_{1}\left( a_{2}^{2}+a_{3}^{2}\right) + 2F_{2}a_{1}^{2}%
\right] \frac{\lambda ^{2}}{a_{2}^{3}a_{3}^{3}} \nonumber \\
&&+G_{2}\left( \frac{lm}{n}%
a_{2}^{4}+\frac{ln}{m}a_{3}^{4}\right) \frac{\lambda }{a_{2}^{3}a_{3}^{3}},
\end{eqnarray}
\begin{eqnarray}
T_{14} &=&-m^2 \left\{\frac{1}{2}G_1\left(a_1^2+a_3^2 \right) + F_2 a_2^2 \right\} \frac{\lambda}{a_1^3a_3^2} \nonumber \\
&&- n^2 \left\{\frac{1}{2}G_1\left(a_1^2+a_2^2 \right) + F_2 a_3^2 \right\} \frac{\lambda}{a_1^2a_2^3}   \nonumber\\
&& - \frac{m^2 n^2}{l^2} \left\{F_1 a_1^2 + F_2 \left(a_2^2+a_3^2 \right) \right\} \frac{\lambda}{a_1^5}   \nonumber\\
&& - \frac{1}{2}G_2\left( \frac{l n}{m} \frac{a_1}{a_2^2} + \frac{l m}{n} \frac{a_1}{a_3^2} \right) - \frac{1}{2} \frac{m n}{l} \frac{F_3}{a_1}, 
\end{eqnarray}%
\begin{eqnarray}
T_{24} &=&- l^2 \left\{\frac{1}{2}G_1\left(a_2^2+a_3^2 \right) + F_2 a_1^2 \right\} \frac{\lambda}{a_2^3 a_3^2} \nonumber\\
&& - n^2 \left\{\frac{1}{2}G_1\left(a_1^2+a_2^2 \right) + F_2 a_3^2 \right\} \frac{\lambda}{a_1^2 a_2^3} \nonumber \\
&& - \frac{l^2 n^2}{m^2} \left\{F_1 a_2^2 + F_2 \left(a_1^2+a_3^2 \right) \right\} \frac{\lambda}{a_2^5}  \nonumber \\
&& - \frac{1}{2}G_2\left( \frac{m n}{l} \frac{a_2}{a_1^2} + \frac{l m}{n} \frac{a_2}{a_3^2} \right) - \frac{1}{2} \frac{l n}{m} \frac{F_3}{a_2} , 
\end{eqnarray}%
and
\begin{eqnarray}
T_{34} &=&- l^2 \left\{\frac{1}{2}G_1\left(a_2^2+a_3^2 \right) + F_2 a_1^2 \right\} \frac{\lambda}{a_2^2 a_3^3} \nonumber\\
&& - m^2 \left\{\frac{1}{2}G_1\left(a_1^2+a_3^2 \right) + F_2 a_2^2 \right\} \frac{\lambda}{a_1^2 a_3^3} \nonumber \\
&& - \frac{l^2 m^2}{n^2} \left\{F_1 a_3^2 + F_2 \left(a_1^2+a_2^2 \right) \right\} \frac{\lambda}{a_3^5}  \nonumber \\
&& - \frac{1}{2}G_2\left( \frac{m n}{l} \frac{a_3}{a_1^2} + \frac{l n}{m} \frac{a_3}{a_2^2} \right) - \frac{1}{2} \frac{l m}{n} \frac{F_3}{a_3} , \label{eqk34}
\end{eqnarray}
with $f_{0}$ defined by equation (\ref{eq46}), where in the case under consideration the parameters $l$, $m$, and $n$ are defined by equations (\ref{defl}) to (\ref{defn}), respectively, and 
\begin{equation}
F_{1}=\frac{\left(3-lmn\right)lmn -\left(lmn\right)^{1/3}-1}{%
(1-lmn)^3\left(lmn\right)^{1/3}}, 
\end{equation}
\begin{equation}
F_{2}=\frac{1-\left(lmn\right)^{1/3}}{(1-lmn)^3\left(lmn\right) ^{1/3}%
}, 
\end{equation}
\begin{equation}
F_{3}=\frac{\left(2lmn-3\right)lmn+\left(lmn\right)^{1/3}}{\left(1-lmn\right)^2\left(
lmn\right) ^{1/3}}, 
\end{equation}
\begin{equation}
G_{1}=\frac{1+lmn-2\left(lmn\right)^{1/3}}{\left(1-lmn\right)^3\left(lmn\right)^{1/3}}%
, 
\end{equation}
and 
\begin{equation}
G_{2}=\frac{\left(lmn\right)^{1/3}-lmn}{\left(1-lmn\right)^2\left(lmn\right)^{1/3}} . 
\end{equation}

 On the external boundary $\lambda \rightarrow 0$, $l \rightarrow 1$, $m \rightarrow 1$, $n \rightarrow 1$, $f_{0} \rightarrow \frac{5}{3}$, and then $k_{11} \rightarrow \frac{5}{3}$, $k_{22} \rightarrow \frac{5}{3}$, $k_{33} \rightarrow \frac{5}{3}$, $k_{12} \rightarrow 0$, $k_{13} \rightarrow 0$, and $k_{23} \rightarrow 0$. So, the infinitely thin focaloid is placed on the external boundary as $M \rightarrow dM$, $l \rightarrow 1$, $m \rightarrow 1$, $n \rightarrow 1$, which implies $\dot{\lambda} \rightarrow 0$. Assuming the additional conditions $\Lambda_i \rightarrow 0$ and ${\dot \Lambda}_i \rightarrow 0$ ({\it i.e.} null vorticity on the boundary or within an infinitely thin focaloid),  equation (\ref{eq75}) reduces to equation (\ref{9a23}) valid for an infinitely thin homoeoid of infinitesimal mass $dM$. 
 
 Also, making $a_1=a_2=a_3=a$, $A_1=A_2=A_3=A$, $l=n=m$, $ \dot{a}_1=\dot{a}_2=\dot{a}_3=\dot{a}$, $ \dot{A}_1=\dot{A}_2=\dot{A}_3=\dot{A}$, yield $\dot{l}=\dot{n}=\dot{m}$,  and so we have $\lambda=a^2 - A^2$, and $\dot{\lambda} =  2 a \dot{a} - 2 A \dot{A}$. Assuming again the additional conditions $\Lambda_i \rightarrow 0$ and ${\dot \Lambda_i} \rightarrow 0$ ({\it  i. e.} null vorticity on the boundary or within the ellipsoidal shell) we verify that equations (\ref{esfv}) and (\ref{K2}) hold also for focaloids in the spherical limit, because the proper summation of the parameters $T_{ij}$, as defined by equations (\ref{eqk11}) to (\ref{eqk34}), converge to the parameters given by equations (\ref{tesf11}), (\ref{tesf12}) and (\ref{t22esf}) in this limit.

\section{Conclusions}\label{sec:Sec6}

This work presents a model description of the kinematical aspects of rotating, perfect, homogeneous fluid ellipsoids and ellipsoidal shells. The equation of continuity is solved allowing an analytical description of the
irrotational velocity field within the system.  Finally, the kinetic energy is expressed analytically for each
specific mass distribution. 

 Nonlinear velocity fields are obtained as solutions of the equation of continuity applied to the fluid motion. We have shown that the ellipsoidal boundaries of the compressible shells are preserved all the time during the internal motion of the fluid content. In addition, it is provided an analytical method to determine the kinetic energy of an inhomogeneous ellipsoidal configuration when it is divided into homogeneous subsystems with specific values of mass, semi-axes, vorticity and angular velocity.

 It is worth mentioning that the procedure described in the present work is not complete, since there are unknown quantities, namely the velocity of outer top axes, $\dot{a}_1$, $\dot{a}_2$, and $\dot{a}_3$ and of the inner top axes, $\dot{A}_1$, $\dot{A}_2$, and $\dot{A}_3$ of the boundary surfaces of the ellipsoidal configurations. Of course, the complete description of the dynamics of the system requires the inclusion of the effects of the gravitation and of the stress tensor, in order to describe the dynamics of the system. 
 
 In view of this, the equations of motion for the top axes must be provided in order to obtain the velocities and the values of the top axes at each instant of time. In the absence of dissipative forces and considering only the gravitational interaction, the virial theorem written in the form of Lagrange's identity is given by \citep{Collins}
\begin{equation}
\frac{1}{2} \frac{d^2I}{dt^2} = 2 T + W,
\end{equation}
where $I$ is the moment of inertia (time dependent) about the origin of the coordinate system, $T$ is the total kinetic energy, including the rotational energy, and $W$ is the total gravitational energy of the system. 
 
 The equations of motion governing the time evolution of the top axes can be obtained, for instance, from the Lagrangian function of the system, $L = T-W$. The resulting equations of motion can be applied together with the virial equation in order to constrain the properties of large-scale celestial bodies in both rigid and differential rotation.
 
  According to the results presented in this work, any homogeneous or even heterogeneous mass distribution that has an ellipsoidal shape can be stratified into any convenient number of homogeneous ellipsoidal shells surrounding a central ellipsoid. Hence, the several procedures presented in the work  can be applied in the computation of the kinetic energy of both the central ellipsoid and each shell. The total kinetic energy is obtained by summing the kinetic energy of the central ellipsoid and each shell, as in equation (\ref{82}). The procedures discussed here can be applied in the both cases of rigid rotation, where the central ellipsoid and the shells have the same angular velocity; and differential rotation (the central ellipsoid and the shells rotate with different angular velocities).
 
  The kinetic energy of the ellipsoidal mass distributions provided in the present work can be useful in the study of the dynamics of some astrophysical systems, in the framework of the models described e. g. in the works given by Saenz and Shapiro \citep{Shapiroa,Shapirob}, Lai et al. \citep{Rasioa,Rasiob}, and Bisnovatyi-Kogan \citep{Kogana,Koganb}. However, this is a subject for  future work.



\end{document}